\documentclass[a4paper,fleqn]{cas-dc}

\usepackage{graphicx}
\usepackage{amsmath}
\usepackage{amssymb}
\usepackage{booktabs}
\usepackage{amsfonts}
\usepackage{multirow}
\usepackage{times}
\usepackage{epsfig}
\usepackage[caption=true]{subfig}
\usepackage{graphicx}
\usepackage{amsmath}
\usepackage{amssymb}
\usepackage{adjustbox}
\usepackage{multirow}
\usepackage{multicol}
\usepackage{booktabs}
\usepackage{url}
\usepackage{array}
\usepackage[accsupp]{axessibility}  
\usepackage{hyperref}
\hypersetup{colorlinks=true}
\hypersetup{
         linkcolor=blue,
         citecolor=blue
}
\usepackage{pifont}
\usepackage[capitalize]{cleveref}

\usepackage[numbers,sort&compress]{natbib}

\usepackage[english]{babel}

\def\tsc#1{\csdef{#1}{\textsc{\lowercase{#1}}\xspace}}
\tsc{WGM}
\tsc{QE}

\newlength\savewidth

\begin{document}
\let\WriteBookmarks\relax
\def\floatpagepagefraction{1}
\def\textpagefraction{.001}

\shorttitle{Adaptive Input-image Normalization for Solving the Mode Collapse Problem in GAN-based X-ray Images}
%
\shortauthors{Saad et~al.}
\title[mode = title]{Adaptive Input-image Normalization for Solving the Mode Collapse Problem in GAN-based X-ray Images}
\tnotemark[1]

\tnotetext[1]{The early version of this paper was published in the EMBC conference: M. M. Saad, M. H. Rehmani and R. O'Reilly, "Addressing the Intra-class Mode Collapse Problem using Adaptive Input Image Normalization in GAN-based X-ray Images," 2022 44th Annual International Conference of the IEEE Engineering in Medicine \& Biology Society (EMBC), Glasgow, Scotland, United Kingdom, 2022, pp. 2049-2052, DOI: \url{10.1109/EMBC48229.2022.9871260.}}

\author[1]{Muhammad Muneeb Saad}[type=editor,
                        orcid=0000-0002-0204-0597
                        ]
\cormark[1]
\ead{muhammad.saad@mycit.ie}
\credit{Conceptualization, Writing- Original draft preparation, Methodology, Experimentation}

\address[1]{Department of Computer Science, Munster Technological University, Cork, Ireland.}

\author[1]
{Mubashir Husain Rehmani}
\credit{Supervision, Review \& Editing}

\author[1]{Ruairi O'Reilly}
\credit{Supervision, Review \& Editing}

\cortext[cor1]{Corresponding author}

\begin{abstract}
Biomedical image datasets can be imbalanced due to the rarity of targeted diseases. Generative Adversarial Networks play a key role in addressing this imbalance by enabling the generation of synthetic images to augment datasets. It is important to generate synthetic images that incorporate a diverse range of features to accurately represent the distribution of features present in the training imagery. Furthermore, the absence of diverse features in synthetic images can degrade the performance of machine learning classifiers. The mode collapse problem impacts Generative Adversarial Networks' capacity to generate diversified images. Mode collapse comes in two varieties: intra-class and inter-class. In this paper, both varieties of the mode collapse problem are investigated, and their subsequent impact on the diversity of synthetic X-ray images is evaluated. This work contributes an empirical demonstration of the benefits of integrating the adaptive input-image normalization with the Deep Convolutional GAN and Auxiliary Classifier GAN to alleviate the mode collapse problems. Synthetically generated images are utilized for data augmentation and training a Vision Transformer model. The classification performance of the model is evaluated using accuracy, recall, and precision scores. Results demonstrate that the DCGAN and the ACGAN with adaptive input-image normalization outperform the DCGAN and ACGAN with un-normalized X-ray images as evidenced by the superior diversity scores and classification scores.
\end{abstract}


\maketitle

\section{Introduction}
Deep learning techniques have shown great success in biomedical image analysis \cite{chen2022recent}. The training of these techniques often requires an ample amount of expert-level annotated image datasets to achieve high accuracy in image recognition \cite{jiang2019task}. However, publicly available biomedical image datasets usually contain a small number of images due to the rarity of diseases and privacy issues of patients \cite{hasani2022artificial}. These biomedical image datasets are limited and insufficient for training deep performant, high-accuracy neural networks \cite{mostapha2019role}. To address this problem, data augmentation is one means of addressing the data limitation problem in biomedical image datasets. Traditional data augmentation techniques include the geometrical transformation of images such as rotation, flipping, zooming, or scaling. These techniques can increase the amount of training image datasets but lack flexibility and are unable to produce diverse image distributions \cite{shen2021mass} \cite{xue2021selective}.

An alternative approach to mitigating the data limitation problem is Generative Adversarial Networks (GANs) \cite{goodfellow2014generative}. GANs have successfully augmented biomedical image datasets by producing synthetic images \cite{karakanis2021lightweight} \cite{wick2021conditional} \cite{shah2022dc} \cite{al2021covid}. A GAN's architecture contains two models: the generator for creating synthetic images and the discriminator for distinguishing synthetic images from real training images. The generator takes a random noisy input, generates raw images as output, and passes them to the discriminator. The discriminator distinguishes synthetic images from real images and back-propagates gradient feedback to the generator. The generator learns the distribution of real images from the discriminator's feedback and endeavors to generate realistically-looking synthetic images \cite{guo2022data}. 

Biomedical images contain vital features that are important to consider when performing image recognition or any other disease-specific computer-aided diagnosis (CAD) task. These features are diverse and are important for training deep learning models in a biomedical setting as opposed to a natural image setting \cite{ma2021understanding}. Such diversified features contain useful information about the disease being detected. Thus, a GAN should generate these diversified salient features while producing synthetic images representative of real biomedical images \cite{abdelhalim2021data}.

In the domain of biomedical imaging, GANs face substantial training challenges such as mode collapse, non-convergence, and instability. These challenges impact the diversity and quality of synthetic images \cite{wang2021generative}. The mode collapse problem is one of the substantial barriers to the generation of diversified images. It can occur in two different forms: intra-class mode collapse, whereby a GAN generates identical synthetic images from distinct input images for a single class, and inter-class mode collapse, whereby a GAN generates identical synthetic images from distinct input images for all classes. It is difficult for a GAN generator to capture all the diverse features in the training images for generating synthetic ones \cite{wick2021conditional}. Both forms of mode collapse directly impact a GAN by degrading its performance in generating diversified synthetic images. Consequently, these less diversified images can degrade the performance of deep learning models that utilize them \cite{liu2023inflating} \cite{torfi2021evaluation}.

Understanding the fundamental characteristics of GANs is crucial for efficient reimplementation and fine-tuning, leading to the generation of high-quality synthetic images, as highlighted in \cite{wang2021generative}. Recent GAN architectures such as StyleGAN2 \cite{karras2019style}, StyleGAN3 \cite{karras2021alias}, and StyleSwin \cite{zhang2022styleswin} have been designed with complex architectures and include additional regularization and normalization techniques. These GAN architectures require optimizing complex objective functions which often necessitates substantial computational power and memory resources for stable training on diverse biomedical images. However, the Deep Convolutional GAN (DCGAN) is designed with a baseline architecture requiring fewer computational resources. DCGAN is a common choice for generating synthetic images in low-resource computational environments \cite{dash2023review}.

Originally, the concept behind GANs was introduced as an unsupervised generative model. There is no class label or image annotation information to be used in baseline GAN variants and no control over the generated data for multi-class datasets \cite{wang2021generative}. Conditional information such as class labels or image annotations can be very useful in controlling the selection of distributions to be generated. The evolution of GANs saw the Conditional GAN (CGAN) introduced \cite{mirza2014conditional}. CGAN architecture has the option to provide class-label information of multi-class datasets to the generator and the discriminator. A CGAN is useful in image generation as it can incorporate conditional feature-based information using image annotations to generate synthetic images that have salient features of interest \cite{golfe2023progleason}. Motivated by this approach, Odena et al. \cite{odena2017conditional} proposed an Auxiliary Classifier GAN (ACGAN), an extension of CGAN. In ACGAN, a modified discriminator is proposed that predicts the class label of input images instead of taking the class label as an input. With this idea, ACGAN can generate more high-quality images than CGAN \cite{karbhari2021generation}.

\begin{table*}[ht]
\centering
\caption{Modified architectures of GAN variants for biomedical image generation}
\begin{footnotesize}
      \begin{tabular}{p{2.5cm}p{0.5cm}p{2cm}p{2cm}p{3cm}p{4cm}}
      \toprule
      \textbf{GAN Variant} & \textbf{Year} & \textbf{Image Type} & \textbf{Intra/Inter Mode Collapse} & \textbf{Application Type} & \textbf{Proposed Solution} \\ \midrule
      MSG-SAGAN \cite{saad2023self} & 2023 & X-ray Images & Intra & Image Synthesis & Self-attention and multi-scale gradients \\
      SPGGAN \cite{abdelhalim2021data} & 2021 & Dermoscopic Images & Intra & Image Synthesis & Self Attention \\
      PGGAN-SSIM \cite{huang2021data} & 2021 & MR Images & Intra & Image Synthesis & SSIM loss \\
      CycleGAN \cite{modanwal2021normalization} & 2021 & MR Images & Intra & Image Translation & Path Discriminator \\
      Improved BAGAN \cite{huang2021enhanced} & 2021 & Cell Images & Intra/Inter & Image Synthesis & Improved Autoencoder \\
      SNSRGAN \cite{xu2020low} & 2020 & X-ray Images & Intra & Image Super Resolution & Spectral Normalization \\
      DCR-AEGAN \cite{segato2020data} & 2020 & MR Images & Intra & Image Synthesis & VAEGAN \\
      SL-StyleGAN \cite{qin2020gan} & 2020 & Dermoscopic Images & Intra & Image Synthesis & Modified Generator \\ \bottomrule
      \end{tabular}
      \end{footnotesize}
      \label{archgan}
\end{table*}

\subsection{Research contributions and objectives}
This work aims to address two main problems in the domain of biomedical image analysis for deep learning researchers. First, alleviate the intra-class and inter-class mode collapse problems in GANs to generate more diversified X-ray images. Second, the augmentation of X-ray images using GAN-based synthetic X-ray images to improve the accuracy of deep learning classifiers.

The research objectives are as follows:
\begin{itemize}
    \item An empirical analysis to evaluate the efficacy of utilizing adaptive input-image normalization (AIIN) to generate more diversified images with DCGAN and ACGAN architectures. Considering parameters are window size, contrast threshold, and batch size.
    \item Evaluating the occurrence of intra-class mode collapse and intra-class diversity using the Multi-scale Structural Similarity Index Measure (MS-SSIM) and Frechet Inception Distance (FID) scores. Furthermore, the inter-class mode collapse problem and diversity of synthetic X-ray images are evaluated by the Inception Score (IS) and FID measures.
\end{itemize}

The research contributions are as follows:
\begin{itemize}
    \item The alleviation of the intra-class mode collapse in DCGAN and the inter-class mode collapse in ACGAN for synthetic X-ray image generation. An adaptive input-image normalization (AIIN) technique is utilized for X-ray images. AIIN is a preprocessing technique that enables GANs to generate highly diversified synthetic X-ray images.
    \item This work contributes to the data augmentation of X-ray images using GAN-based synthetic X-ray imagery. For this purpose, the best GAN-based diversified synthetic images, as evident from MS-SSIM, IS, and FID scores, are used with the real X-ray images to improve the performance of deep learning classifiers.
\end{itemize}

\begin{table}[hbt!]
\vspace{-1em}
\centering
\caption{An overview of preprocessing techniques in GAN-based image generation for biomedical image analysis.}
\begin{footnotesize}
      \begin{tabular}{p{2.5cm}p{0.5cm}p{1cm}p{2.5cm}}
      \toprule
      \textbf{GAN Variant} & \textbf{Year} & \textbf{Image Type} & \textbf{Proposed Solution} \\ \midrule
      DCGAN \cite{saad2022addressing} & 2022 & X-ray & AIIN \\
      CGAN \cite{qadir2021covid} & 2021 & X-ray & Edge Detection \\
      DAPR-Net \cite{huang2020dapr} & 2020 & Retinal & HE \\
      DCGAN/CGAN \cite{verma2020synthetic} & 2020 & Protein & Contrast-limiting HE \\ \bottomrule
      \multicolumn{4}{l}{AIIN: Adaptive Input-image Normalization} \\
      \multicolumn{4}{l}{HE: Histogram Equalization}
      \end{tabular}
      \end{footnotesize}
      \label{pregan}
      \vspace{-2em}
\end{table}

This work is an extension of our prior work based on the following contributions:

\subsection{Our Prior work}
In prior work \cite{saad2022addressing}, AIIN was proposed to address the intra-class mode collapse in DCGAN for generating diversified synthetic X-ray images. AIIN assisted in alleviating the intra-class mode collapse in DCGAN and improved the intra-class diversity of generated images as evidenced by the MS-SSIM and FID scores. A CNN classifier was used to evaluate the performance improvement of augmented datasets using GAN-based generated images. The classification performance including accuracy and specificity was improved using normalized GAN-based generated images. However, the work was limited to the evaluation of the intra-class mode collapse problem. This work extends that approach to the alleviation of the inter-class mode collapse problem for X-ray image generation. Novel contributions and differences from prior work are outlined as follows:
\begin{itemize}
    \item In our prior work, AIIN was integrated with DCGAN to evaluate the intra-class mode collapse problem for healthy X-ray images. The identification of intra-class mode collapse and diversity of generated images were assessed using the MS-SSIM and FID scores. 
    \item This work extends the prior approach of integrating AIIN with DCGAN to alleviate intra-class mode collapse for coronavirus and Pneumonia diseases through chest X-ray images. This work helps to investigate the generalization of the approach used in the prior work.
    \item This work also contributes to integrating AIIN with ACGAN to alleviate the inter-class mode collapse problem for coronavirus and Pneumonia diseases based on binary X-ray image datasets. The diversity of generated images is evaluated using IS, FID, and MS-SSIM scores.
    \item In this work, a state-of-the-art pre-trained Vision Transformer (ViT) based deep learning classifier is implemented for the classification of X-ray images using GAN-based augmented datasets. 
\end{itemize}

\begin{table}[hbt!]
\centering
\caption{GAN-based augmentation approaches for X-ray image analysis}
\begin{footnotesize}
      \begin{tabular}{p{1.1cm}p{0.35cm}p{0.35cm}p{1.5cm}p{1.25cm}p{1.38cm}}
      \toprule
      \textbf{Aug.} & \textbf{Trad.} & \textbf{Year} & \textbf{Classifier} & \textbf{Pre-Aug. Score} & \textbf{Post-Aug. Score} \\ \midrule
      DCGAN \cite{barshooi2022novel} & Yes & 2022 & DenseNet-201 & Acc:77.0 & Acc:98.5 \\
      DCGAN \cite{shah2022dc} & No & 2022 & EfficientNetB4 & AUC:92.0 & AUC:96.0 \\
      WGAN \cite{hussain2022wasserstein} & No & 2022 & Lightweight CNN & Acc:93.04 & Acc:95.34 \\
      CGAN \cite{al2021covid} & No & 2021 & SqueezeNet & No Inf. & Acc:98.86 \\
      ACGAN \cite{karbhari2021generation} & No & 2021 & VGG16-HS & Acc:99.36 & Acc:100.0 \\
      CGAN \cite{karakanis2021lightweight} & No & 2021 & CNN & Acc:96.5 & Acc:98.7 \\
      ACGAN \cite{waheed2020covidgan} & No & 2020 & CNN & Acc:85.0 & Acc:95.0 \\
      CDCGAN \cite{zulkifley2020covid} & No & 2020 & LightCovidNet & Acc:92.37 & Acc:96.97 \\
      \bottomrule
      \multicolumn{6}{l}{Aug: Augmentation; Acc: Accuracy; Inf: Information} \\
      \multicolumn{6}{l}{Trad: Traditional}
      \end{tabular}
      \end{footnotesize}
      \label{datauggan}
\end{table}
The remainder of the paper is organized as follows; Section \ref{sec:related_work} presents the literature review of related work studies. Section \ref{sec:methodology} discusses the proposed methodology for the experiments and evaluation metrics. Section \ref{sec:results} outlines the results. Section \ref{discussion} presents a discussion of the research findings and compares them with state-of-the-art prior approaches. Section \ref{sec:conclusion} concludes this paper.

\section{Related Work}
\label{sec:related_work}
Deep learning researchers have successfully utilized GANs for several applications such as image synthesis, image segmentation, image reconstruction, and object detection in the biomedical imaging domain \cite{singh2021medical} \cite{kazeminia2020gans}. While performing these tasks, GANs face technical challenges such as mode collapse, during training. Several approaches have been proposed to address the mode collapse problem for various GAN architectures. Self-attention mechanism in Progressive Growing of GAN (PGGAN) was proposed to generate more diversified skin lesion images \cite{abdelhalim2021data}. The efficacy of the attention mechanism is discussed, stating that it helps in the coordination of salient features at every location to capture the long-range dependencies in biomedical images. Similarly, Huang et al. \cite{huang2021data} used the SSIM loss in the PGGAN architecture to address the mode collapse problem while generating diversified Magnetic Resonance (MR) images. Modanwal et al. \cite{modanwal2021normalization} utilized 34x34 small patches of images in the discriminator to alleviate the mode collapse in CycleGAN for MR image generation. Small patches help to maintain the structural information of dense tissues during image translation tasks. Contributions such as the improved autoencoder \cite{huang2021enhanced}, spectral normalization \cite{xu2020low}, variational GAN \cite{segato2020data}, and improved style generator \cite{qin2020gan} have been made to alleviate the mode collapse problem in GANs.

Unconditional GANs suffer from intra-class mode collapse while conditional GANs suffer from intra-class as well as inter-class problems. Table \ref{archgan} summarizes the GAN variants with their contributions to specific biomedical imaging modalities. These contributions are application-specific and partially alleviate the mode collapse problem for the given variant of GANs.

Preprocessing techniques have also been integrated into GAN architectures to generate more diverse synthetic images in the biomedical imaging domain. Preprocessing techniques improve the visibility of diversified features in biomedical images. In this context, Saad et al. \cite{saad2022addressing} utilized an adaptive input-image normalization technique that normalizes X-ray image features using contrast histogram equalization. The authors evaluated this normalization approach for the DCGAN architecture to generate diversified X-ray images. Similarly, Qadir et al. \cite{qadir2021covid} used the segmented masks of chest edges with the training X-ray images in CGAN to generate more diversified synthetic images. Histogram equalization techniques have been used in preprocessing stages for Retinal and Protein image synthesis \cite{huang2020dapr} \cite{verma2020synthetic}. Table \ref{pregan} summarizes the preprocessing techniques utilized for GANs. However, it is necessary to investigate the efficacy of preprocessing techniques in improving the intra-class and inter-class diversities of GAN-based generated images via evaluation metrics and classification scores.
 
Previously, augmented datasets using GAN-based synthetic images have improved the performance of deep learning classifiers in biomedical image analysis. Unconditional and conditional GANs, both have performed well in several data augmentation tasks for biomedical (X-ray) image analysis, as summarized in Table \ref{datauggan}. The DCGAN has been utilized to generate X-ray images improving the performance of DenseNet-201 \cite{barshooi2022novel} and EfficientNetB4 \cite{shah2022dc}. Conditional GANs have been utilized to augment the imbalanced X-ray datasets, improving the classification accuracy of SqueezNet \cite{al2021covid}, ResNet34 \cite{xue2021selective}, CNN \cite{karakanis2021lightweight}, and LightCovidNet \cite{zulkifley2020covid}. Previous work discussed that the conditional GAN variants suffer from a mode collapse problem that degrades the diversity of generated images. These less diverse images fail to improve the performance of classifiers trained on these augmented datasets \cite{xue2021selective} \cite{karbhari2021generation} \cite{zulkifley2020covid}.

\begin{figure*}
    \centering
    \includegraphics[width=1\textwidth]{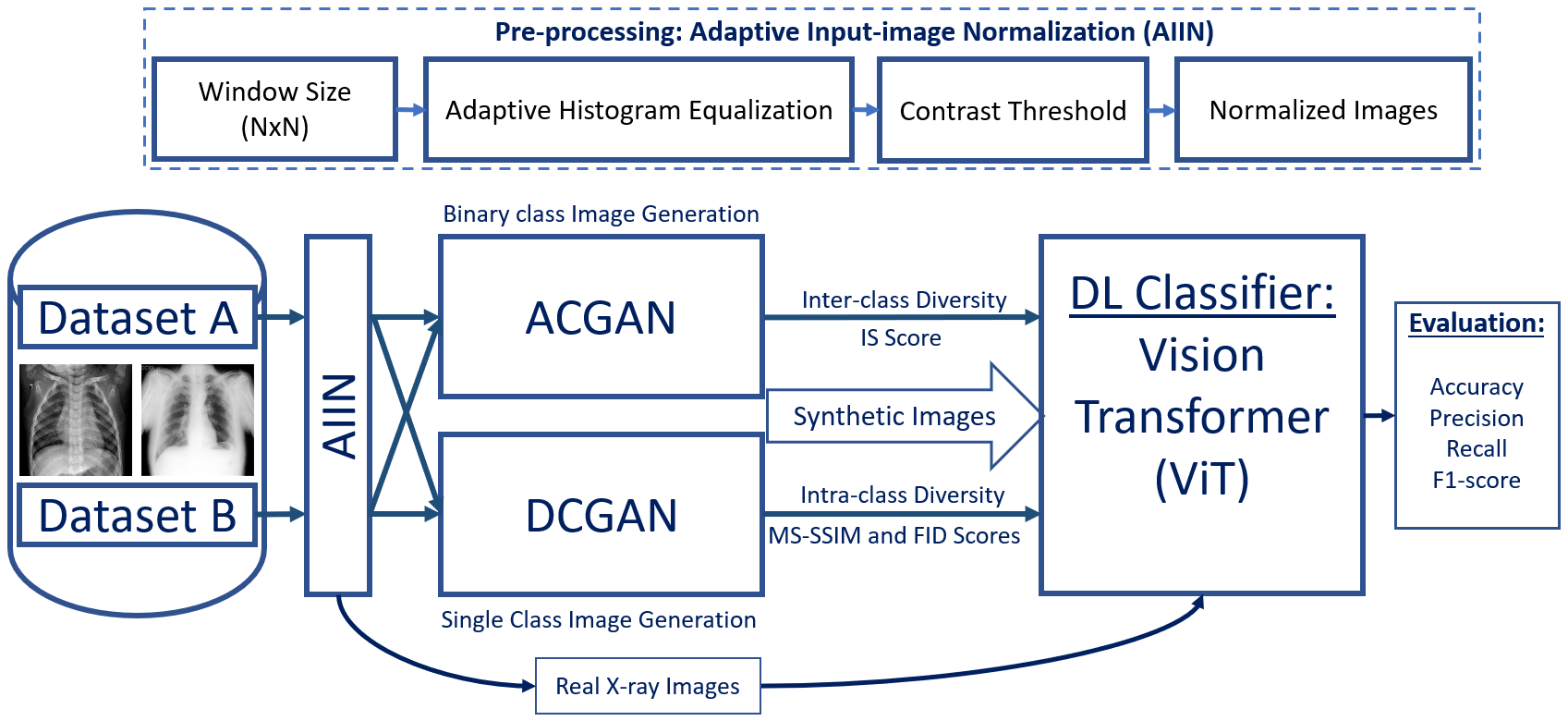}
    \caption{Proposed methodology for data augmentation of chest X-ray datasets with GAN-based generated images.}
    \label{fig:flowdiagram}
    \vspace{-1em}
\end{figure*}

\begin{figure*}[hbt!]
    \vspace{-1em}
    \centering
    \includegraphics[width=1\textwidth]{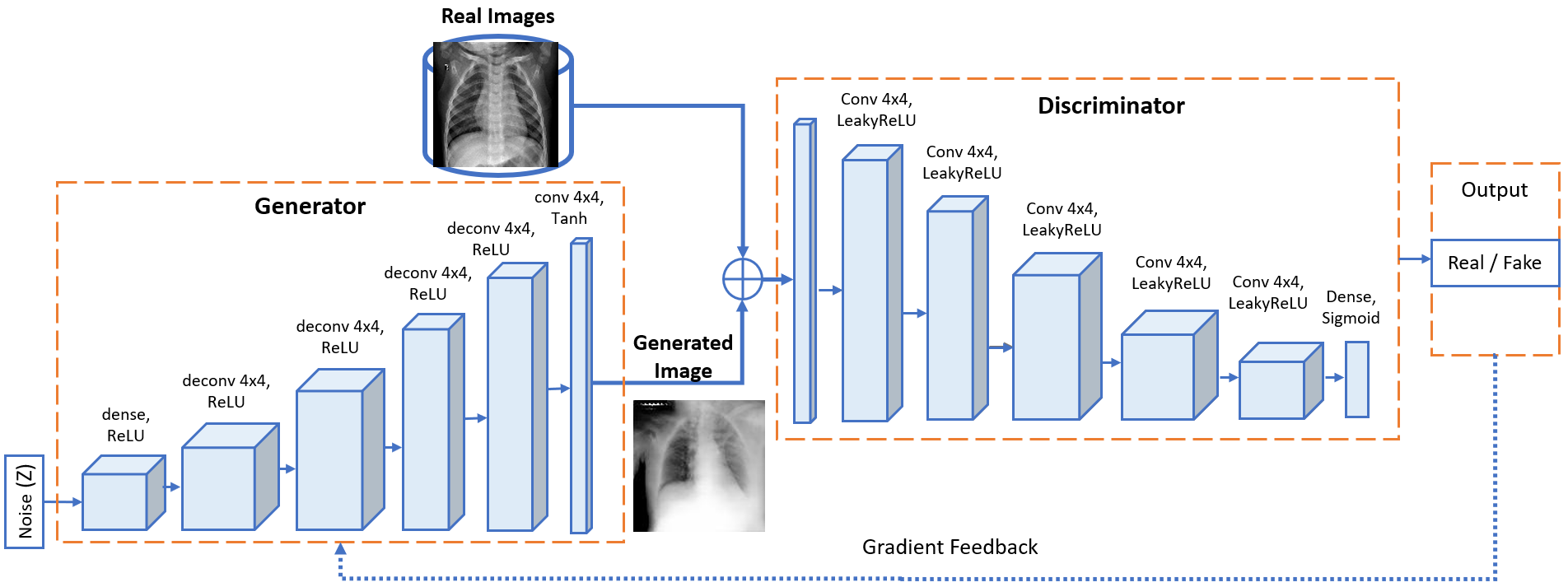}
    \caption{Block diagram of DCGAN architecture \cite{saad2022addressing}.}
    \label{fig:dcgan_arch}
    \vspace{-1em}
\end{figure*}

\begin{figure*}[hbt!]
    \centering
    \includegraphics[width=1\textwidth]{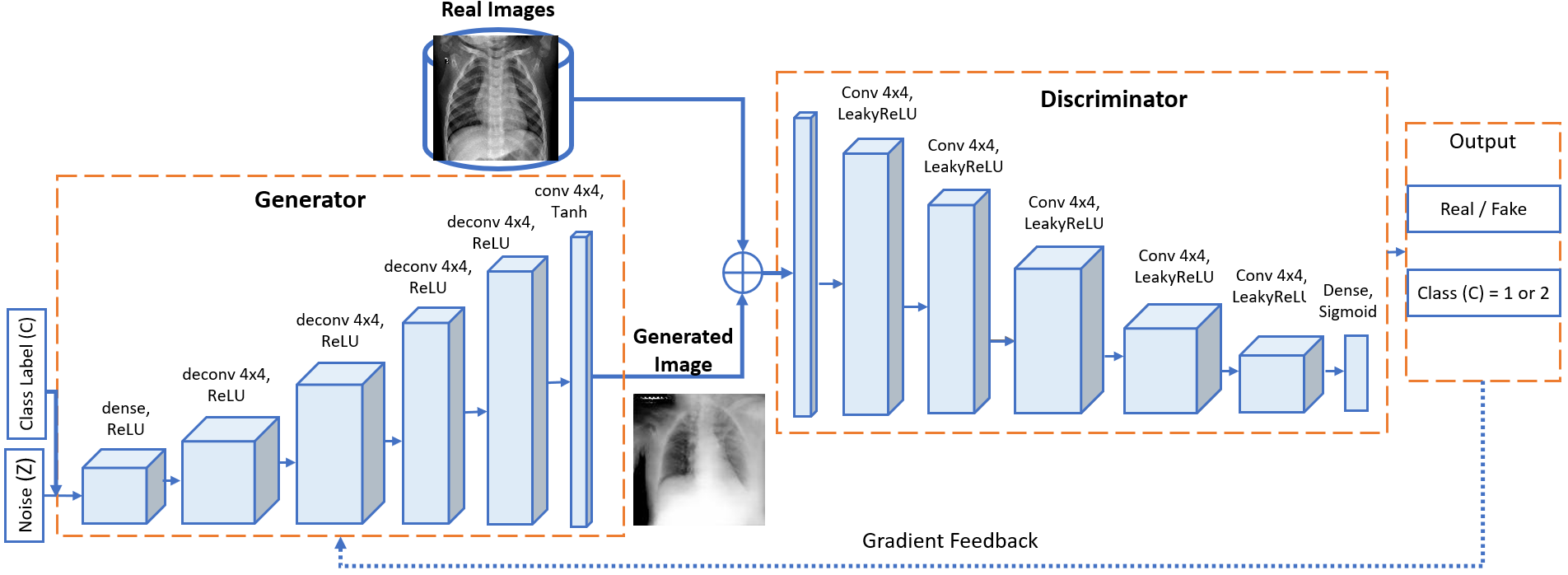}
    \caption{Block diagram of ACGAN architecture.}
    \label{fig:acgan_arch}
\end{figure*}

\begin{figure*}[hbt!]
  \centering
\subfloat[Real and DCGAN generated synthetic COVID-19 and Pneumonia samples from un-normalized and AIIN normalized X-ray images.]{%
  \includegraphics[clip,width=0.9\textwidth]{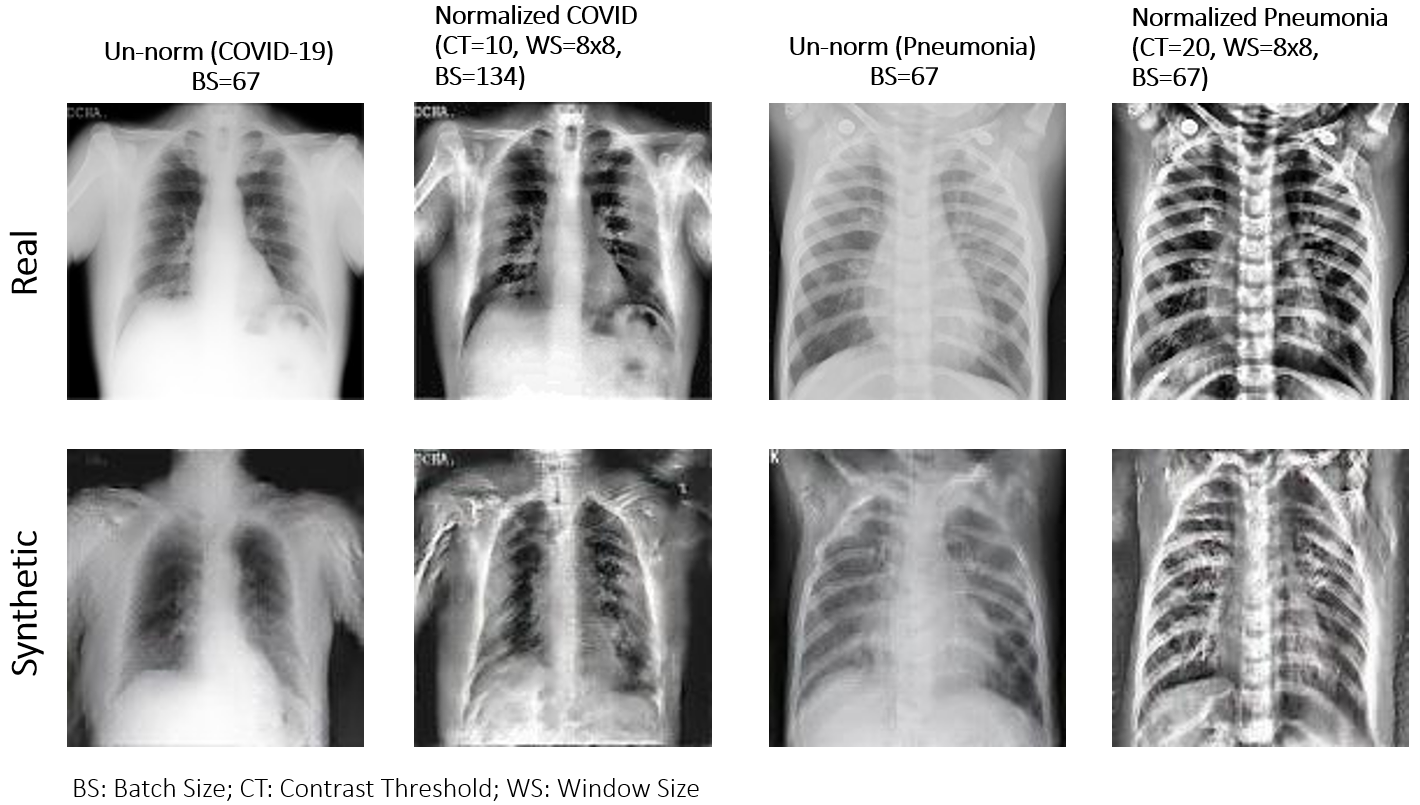}%
}

\subfloat[Real and ACGAN generated synthetic COVID-19 and Pneumonia samples from un-normalized and AIIN normalized X-ray images.]{%
  \includegraphics[clip,width=0.9\textwidth]{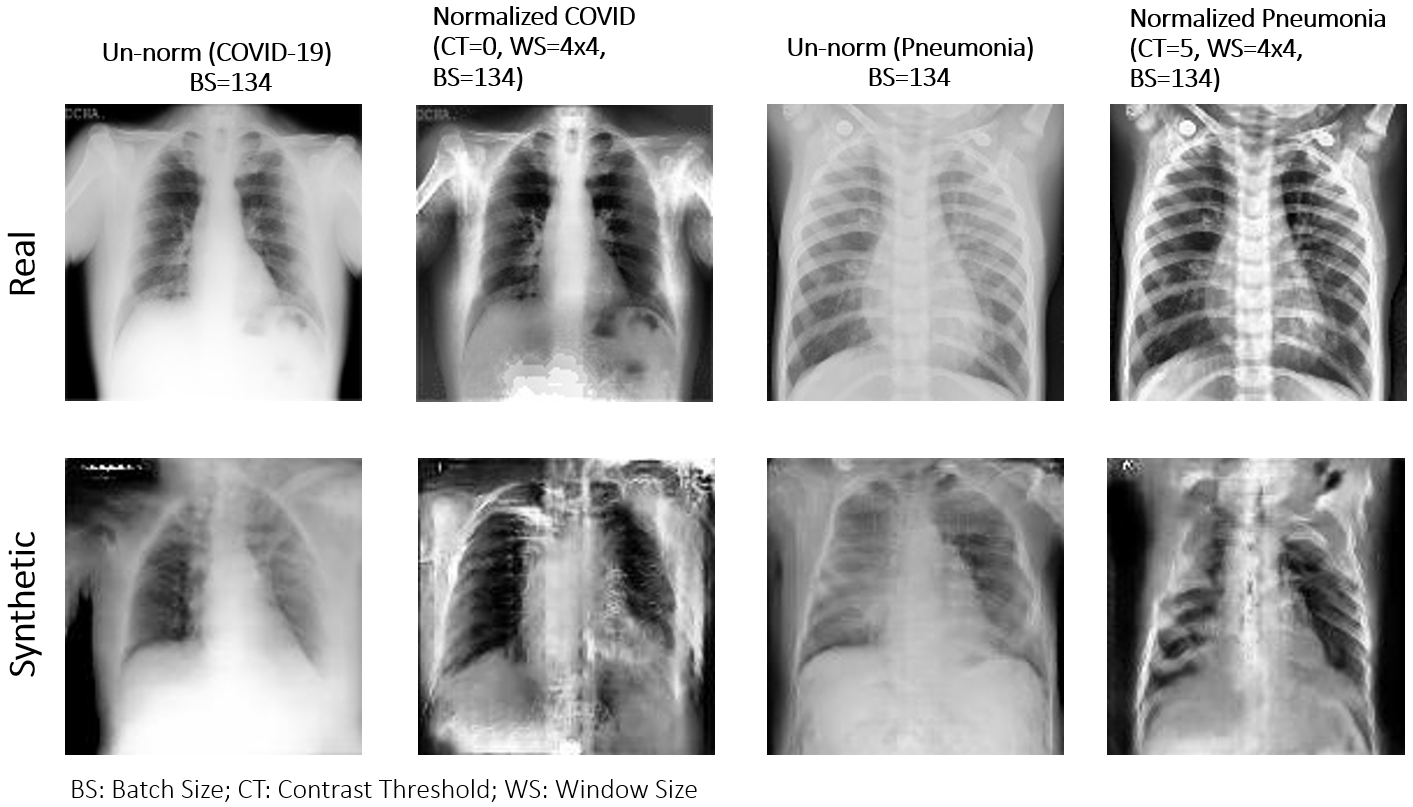}%
}
  \caption{COVID-19 and Pneumonia real and synthetic samples from the un-normalized and AIIN normalized X-ray images.}
  \label{Fig_samples}
\end{figure*}

\begin{figure*}[hbt!]
  \vspace{-1em}
  \centering
\subfloat[MS-SSIM scores enabling an assessment of the occurrence of intra-class mode collapse.]{%
  \includegraphics[clip,width=0.95\textwidth,height=5.5cm]{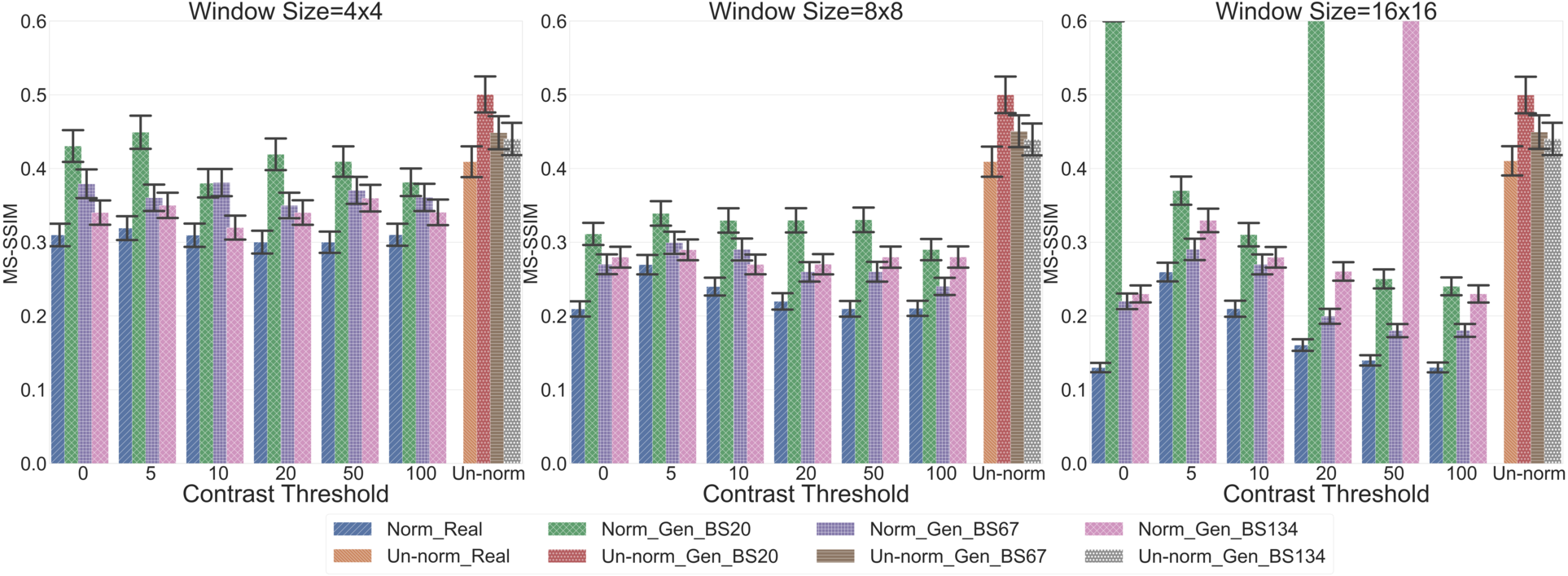}%
}

\subfloat[FID scores enabling an assessment of the level of intra-class diversity.]{%
  \includegraphics[clip,width=0.95\textwidth,height=5.5cm]{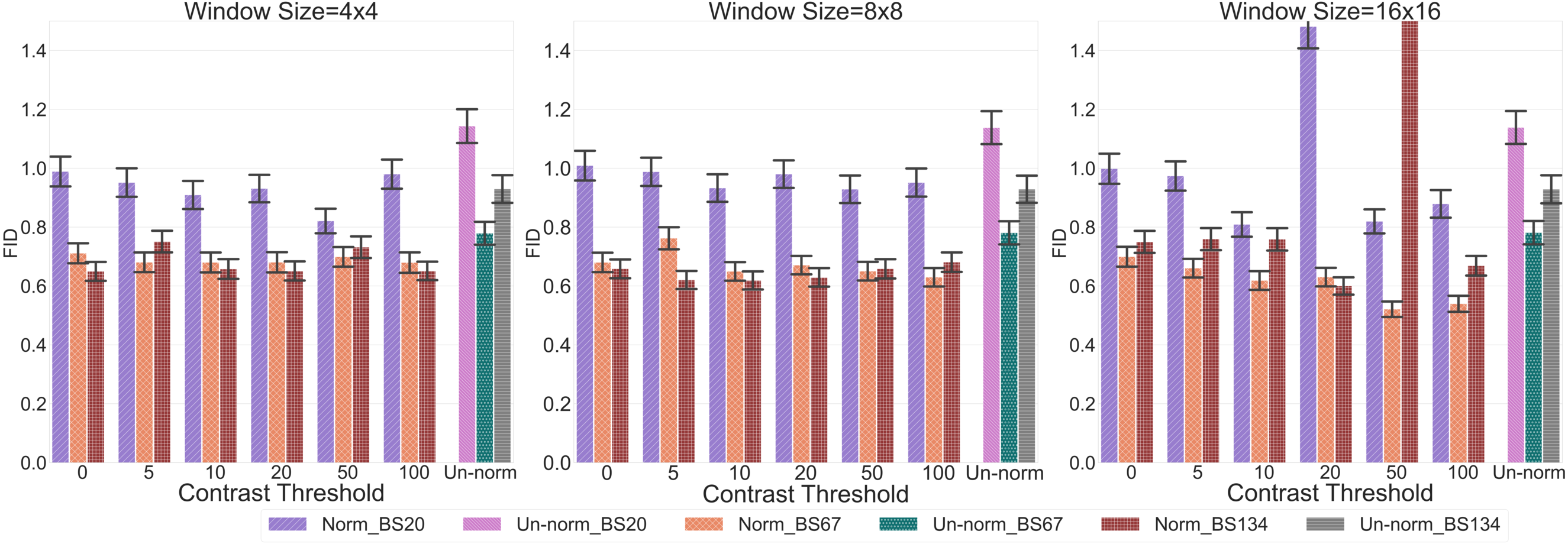}%
}
  \caption{The MS-SSIM and FID scores of DCGAN for the un-normalized and AIIN normalized COVID-19 X-ray images.}
  \label{MS-SSIM_FID_covid}
  \vspace{-1em}
\end{figure*}
\begin{figure*}[hbt!]
    \vspace{-1em}
    \centering
\subfloat[MS-SSIM scores enabling an assessment of the occurrence of intra-class mode collapse.]{%
  \includegraphics[clip,width=0.95\textwidth,height=5.5cm]{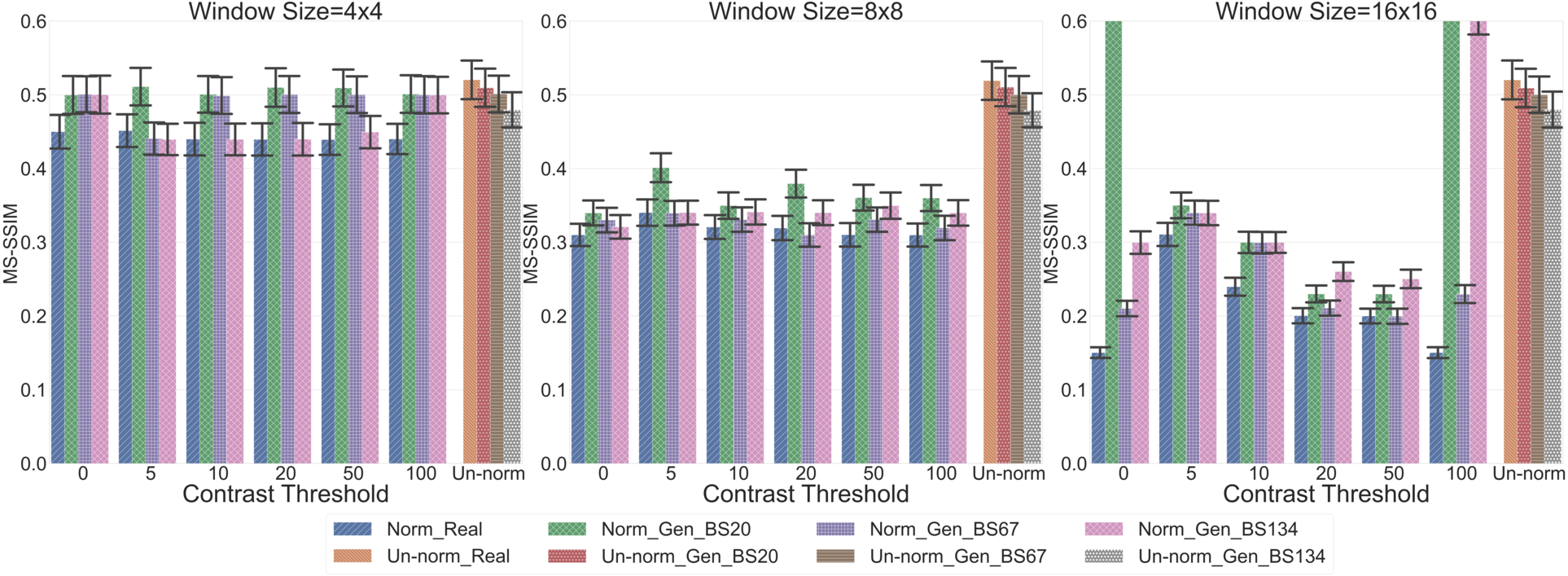}%
}

\subfloat[FID scores enabling an assessment of the level of intra-class diversity.]{%
  \includegraphics[clip,width=0.95\textwidth,height=5.5cm]{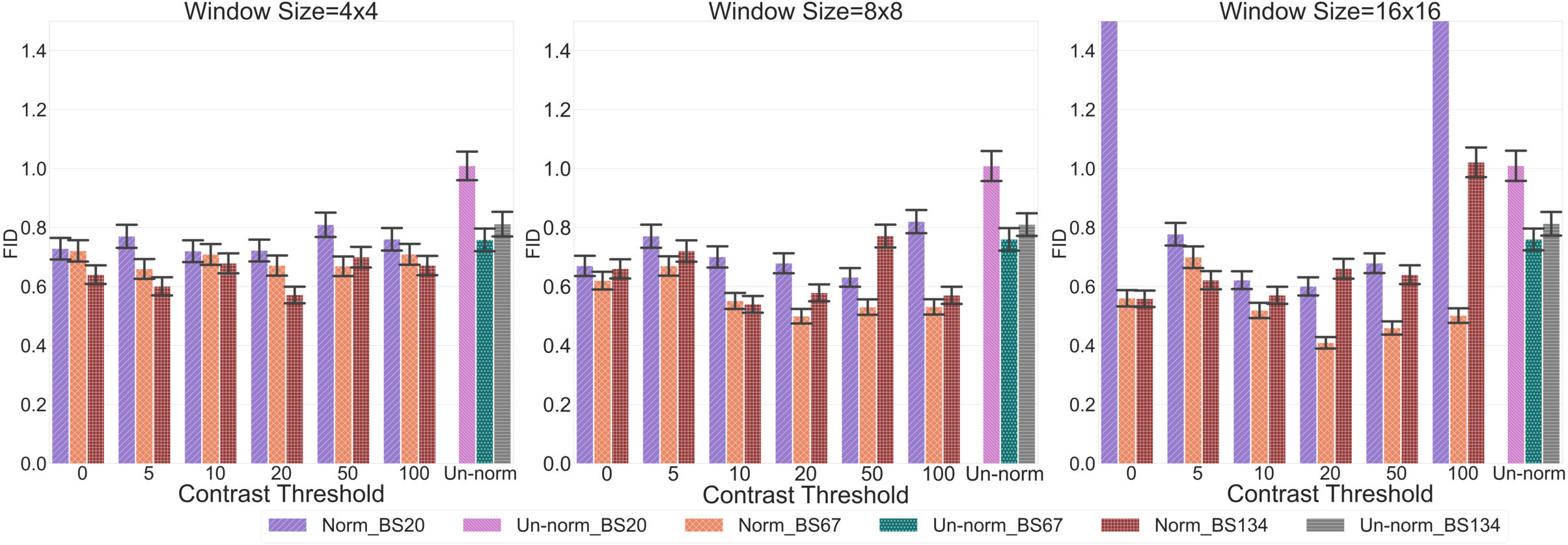}%
}
  \caption{The MS-SSIM and FID scores of DCGAN for the un-normalized and AIIN normalized Pneumonia X-ray images.}
  \label{MS-SSIM_FID_pneu}
  \vspace{-1em}
\end{figure*}

\section{Methods}
\label{sec:methodology}
The proposed methodology is depicted in Fig. \ref{fig:flowdiagram}. Synthetic images of DCGAN and ACGAN are used for the augmentation of imbalanced datasets A and B. AIIN is utilized with both GAN models for the improved diversified generation of images.
\subsection{Datasets}
In this work, a publicly available dataset \cite{rahman2021exploring} is utilized as the source dataset. The source dataset contains chest X-ray images of 3616 coronavirus disease (COVID-19), 10192 healthy, 6012 lung opacity, and 1340 Pneumonia as accessed from \cite{source_xray_data}. The source dataset was compiled by selecting X-ray images from other publicly available datasets. There was no information provided on patients related to the images. It is mentioned that COVID-19, lung opacity, and a few normal images were acquired from 15000 patients \cite{rahman2021exploring}. However, the patient-related image information was missing \cite{rahman2021exploring}. The 1340 COVID-19, 1340 Pneumonia, and 2680 healthy chest X-ray images for the training dataset, while 268 COVID-19, 268 Pneumonia, and 536 healthy chest X-ray images for the test dataset were selected from the source dataset to replicate the experimental setup for training GANs in \cite{saad2022addressing}. Selected images are further distributed into two datasets A and B as detailed in Table \ref{dataset}. Both datasets are imbalanced with the COVID-19 X-ray images being the minority class in dataset A while Pneumonia X-ray images being the minority class in the dataset B. The augmentation of these minority class images is required to balance the dataset. All images were resized into 128x128 as detailed in \cite{saad2022addressing}. Samples of COVID-19 and Pneumonia images are depicted in Fig. \ref{Fig_samples}.
\begin{table}[hbt!]
\centering
\caption{\textbf{Distribution of chest X-ray images \cite{rahman2021exploring}.}}
\begin{tabular}{p{0.3cm}p{0.5cm}p{0.8cm}p{1.3cm}p{0.5cm}p{0.8cm}p{1cm}}
\toprule 
& \multicolumn{3}{c|}{\textbf{Dataset A}} & \multicolumn{3}{|c}{\textbf{Dataset B}} \\
& Total Img. & Healthy & COVID-19 & Total Img. & Healthy & Pneumonia \\
\midrule

Train & 4020 & 2680 & 1340 & 4020 & 2680 & 1340 \\
Test & 804 & 536 & 268 & 804 & 536 & 268 \\
\bottomrule
\multicolumn{7}{l}{Img: Images} \\
\end{tabular}
\label{dataset}
\end{table}

\subsection{Data Preprocessing}
The input images are normalized using AIIN for both GAN architectures. AIIN uses contrast-based histogram equalization to normalize the images \cite{clahe1994contrast}. In this process, contrast, one of the morphological features is used to highlight the diverse features of the images. An X-ray image is segmented into non-overlapping NxN windows. The histogram of each window is normalized locally with a series of contrast threshold values. A contrast threshold value helps to prevent the amplification of noise. After normalization, all the individual windows are joined together using bilinear interpolation. This technique helps to highlight the diverse features of chest X-ray images so that the GANs can generate diversified images effectively. This work adopts AIIN with window sizes 4x4, 8x8, and 16x16 for X-ray images. Contrast threshold values (0, 5, 10, 20, 50, 100) were selected to normalize the images \cite{saad2022addressing}. In our prior work \cite{saad2022addressing}, alternate preprocessing methods such as Gaussian and Mean filtering techniques were also applied. It was found that the effect was insignificant for GAN-based X-ray image synthesis because image features were suppressed to degrade the feature information.  
\subsection{GAN Architectures}
The DCGAN and ACGAN architectures have been implemented for Datasets A and B. The DCGAN is implemented for a single class (COVID-19 X-ray images from dataset A and Pneumonia X-ray images from dataset B) while the ACGAN is implemented for a binary class X-ray image dataset. In this work, the DCGAN has identical architecture settings as depicted in Fig. \ref{fig:dcgan_arch}. The ACGAN architecture has been adopted from \cite{waheed2020covidgan} with some modifications. The detailed layered architecture of ACGAN is depicted in Fig. \ref{fig:acgan_arch}. In the ACGAN architecture, a latent random input $z$ of 2000 with the associated class labels $c$ is used. The discriminator of the ACGAN architecture is composed of five convolutional layers. Each convolutional layer is followed by batch normalization with a momentum of 0.8, Leaky ReLU with a slope of 0.2, and a dropout with 0.5 probability layers to downsample the images. Similarly, the generator of ACGAN contains an input layer, an embedding layer of 50 dimensions, and four convolutional-transpose layers to upsample the images. The batch normalization follows each convolutional-transpose layer with a momentum of 0.8 and the ReLU activation layers except for the last layer which only uses the tanh activation function. Both GAN architectures are trained for 500 epochs to enable the convergence of both generator and discriminator models in the GAN. The un-normalized and AIIN-normalized X-ray images from Datasets A and B are used for training the DCGAN and ACGAN architectures. Three different batch sizes (i.e., 20, 67, and 134) are used for training. The choice of batch size was made based on the number of images used for training the GANs. These batch sizes indicate a number divisible by the total number of images so that all images can be used for training the GANs as suggested in \cite{saad2022addressing}.

\subsection{Mode Collapse and Diversity of Synthetic Images}
The occurrence of intra-class mode collapse is identified using MS-SSIM and inter-class mode collapse using IS scores. The diversity of generated synthetic images is assessed by the FID score. These metrics enable the evaluation of a GAN's capacity to generate diversified synthetic images.
\subsubsection{Mode Collapse Problem}
MS-SSIM score is a widely used metric to measure the intra-class mode collapse problem in GANs by assessing the intra-class diversity of synthetic images \cite{odena2017conditional}. MS-SSIM computes the similarity score between two images randomly selected from a dataset. MS-SSIM scores of real and synthetic image datasets are compared to find the difference in diverse features of both datasets. A higher MS-SSIM score of the synthetic dataset indicates the occurrence of mode collapse in GANs. In this work, 670 image pairs are randomly selected for COVID-19 and Pneumonia images from datasets A and B.
Let's say x and y are two image samples. MS-SSIM is computed between these two image samples as defined in Eq. \eqref{Eq. ms-ssim} \cite{borji2019pros}.
\begin{equation}
\resizebox{0.9\hsize}{!}{$%
\operatorname{MS}-\operatorname{SSIM}(x, y)=I_{M}(x, y)^{\alpha_{M}} \prod_{j=1}^{M} C_{j}(x, y)^{\beta_{j}} S_{j}(x, y)^{\gamma_{j}}\label{Eq. ms-ssim}
$}%
\end{equation}
Contrast (C) and structural (S) features of images are computed at scale $j$ as denoted in Eq. \eqref{Eq. ms-ssim}. Luminance is calculated at the coarsest scale (M). The $\alpha$, $\beta$, and $\gamma$ are the weight parameters as detailed in \cite{wang2003multiscale}.

The inter-class mode collapse is measured by the IS score. IS computes the Kullback–Leibler divergence between the class conditional probability $p(y \mid \mathbf{x})$ of each generated image and the marginal probability $p(y)$ calculated from a group of images generated from all classes. IS measures the lowest score as 1 and higher as equal to the number of classes. It estimates the upper bound as the higher score shows that the model can generate diversified and high-quality images. IS shows an acceptable correlation with the diversity and quality of generated images and uses a pre-trained Inception-Net for the assessment of generated images. It is defined in Eq. \eqref{Eq. IS}:
\begin{equation}
\exp \left(\mathbb{E}_{\mathbf{x}}[\mathbb{K} \mathbf{L}(p(\mathbf{y} \mid \mathbf{x}) \| p(\mathrm{y}))]\right)=\exp \left(H(y)-\mathbb{E}_{\mathbf{x}}[H(y \mid \mathbf{x})]\right)\label{Eq. IS}
\end{equation}
In Eq. \eqref{Eq. IS}, class conditional probability, marginal probability, and entropy are denoted by $p(y \mid \mathbf{x})$, $p(y)$, and $H(x)$ for image samples $x$.

\subsubsection{Diversity of Synthetic Images}
The intra-class and inter-class diversity of synthetic images is assessed by the FID and IS scores. IS estimates the diversity of synthetic images generated from multiple classes. It has some limitations as it only relies on generated synthetic images without comparing them to real images. Whereas, FID estimates the distance between feature activations of real and feature activations of synthetic images \cite{miyato2018cgans} from single class images. FID score ranges between 0.0 and $+\infty$ while a lower score indicates a higher diversity of synthetic images as compared to the real images. In this work, 1340 samples were selected from real images and 1340 samples from synthetic images for measuring FID scores. IS score is measured using 1000 images from each class of generated images. 
\subsection{Assessing the Utility of Synthetic Images}
GAN-based synthetic generated X-ray images are used for the augmentation of minority classes in datasets A and B. Dataset A is augmented with 1340 synthetic COVID-19 X-ray images while dataset B is augmented with 1340 synthetic Pneumonia X-ray images. A state-of-the-art Vision Transformer (Vit) model \cite{vit} pre-trained on ImageNet is implemented for the classification of COVID-19 and Pneumonia chest X-ray images.

To augment the imbalance datasets A and B, GAN-based synthetic generated images with higher diversified scores of MS-SSIM, IS, and FID scores are selected from each window size. For training the ViT classifier, all the images were rescaled to 224x224 size. A batch size of 16 was used to train the ViT classifier. Results are assessed with accuracy score, precision, recall, and F1 scores. Classification scores are compared with the un-normalized synthetic X-ray images for all instances as detailed in Table \ref{tabclasscompare}.

\section{Results}
\label{sec:results}
The MS-SSIM and FID scores of DCGAN for the un-normalized and AIIN normalized X-ray images are depicted in Fig. \ref{MS-SSIM_FID_covid} and Fig. \ref{MS-SSIM_FID_pneu}. The IS and FID scores of ACGAN for the un-normalized and AIIN normalized X-ray images are depicted in Fig. \ref{IS_FID_covid} and Fig. \ref{IS_FID_pneu}. Figure \ref{MS-SSIM_FID_covid} shows the MS-SSIM and FID scores while Fig. \ref{IS_FID_covid} shows the IS and FID scores of COVID-19 real and synthetic generated X-ray images. Similarly, Fig. \ref{MS-SSIM_FID_pneu} depicts the MS-SSIM and FID scores while Fig. \ref{IS_FID_pneu} depicts the IS and FID scores of Pneumonia real and generated X-ray images.

Synthetic X-ray images of GAN-based variants with higher diversified scores have been utilized to balance the actual datasets. A ViT classifier has trained on these datasets. Results of ViT show that the classification results vary across different disease X-ray images in terms of accuracy, recall, precision, and F1 scores as indicated in Table \ref{tabclasscompare}. AIIN-normalized images have degraded the ViT classification scores in augmenting the imbalanced datasets. Contrarily, ViT results are improved for Pneumonia X-ray images using AIIN-normalized augmented datasets.

\section{Discussion}
\label{discussion}
This paper aims to investigate the use of AIIN, a preprocessing technique that can enable GAN-based models to produce diversified images in the biomedical imaging domain, for intra-class and inter-class diversities of GAN-based synthetic images. AIIN is applied to the input images as a pre-processing in the proposed AIIN-DCGAN and AIIN-ACGAN. AIIN is applied to the input images as a pre-processing in the proposed AIIN-DCGAN and AIIN-ACGAN that enhances the prominence of diverse input features using contrast-based histogram equalization. The diverse features include the shape and texture of body parts in a biomedical image. In the context of chest X-ray images, these features include the spine, heart, and lungs with their visual signatures like ribs, aortic arch, and distinct curvature of the lower lungs. These features are overlooked by the discriminator during the training of GANs, enabling the generator to produce severely mode-collapsed images. The training of GANs results in the mode collapse problem when it is stuck on generating irregular shapes of lungs repetitively without focusing on minor details of diverse features. To address this problem, contrast-based histogram equalization in AIIN is used to normalize the contrast of diverse features of images \cite{clahe1994contrast} \cite{adaptivehisto}. By this approach, diverse features in X-ray images are highlighted and become prominent to the discriminator in the training GANs. The discriminator learns these features more accurately and provides constructive gradient feedback to the generator using normalized X-ray images. Consequently, the generator is forced to produce more diversified images, alleviating the mode collapse problem.

\begin{figure*}[hbt!]
  \vspace{-1em}
  \centering
\subfloat[DCGAN Training behavior on COVID-19 X-ray images over epochs. The MS-SSIM scores of real and generated images are shown in the same colors.]{%
  \includegraphics[clip,width=1\textwidth]{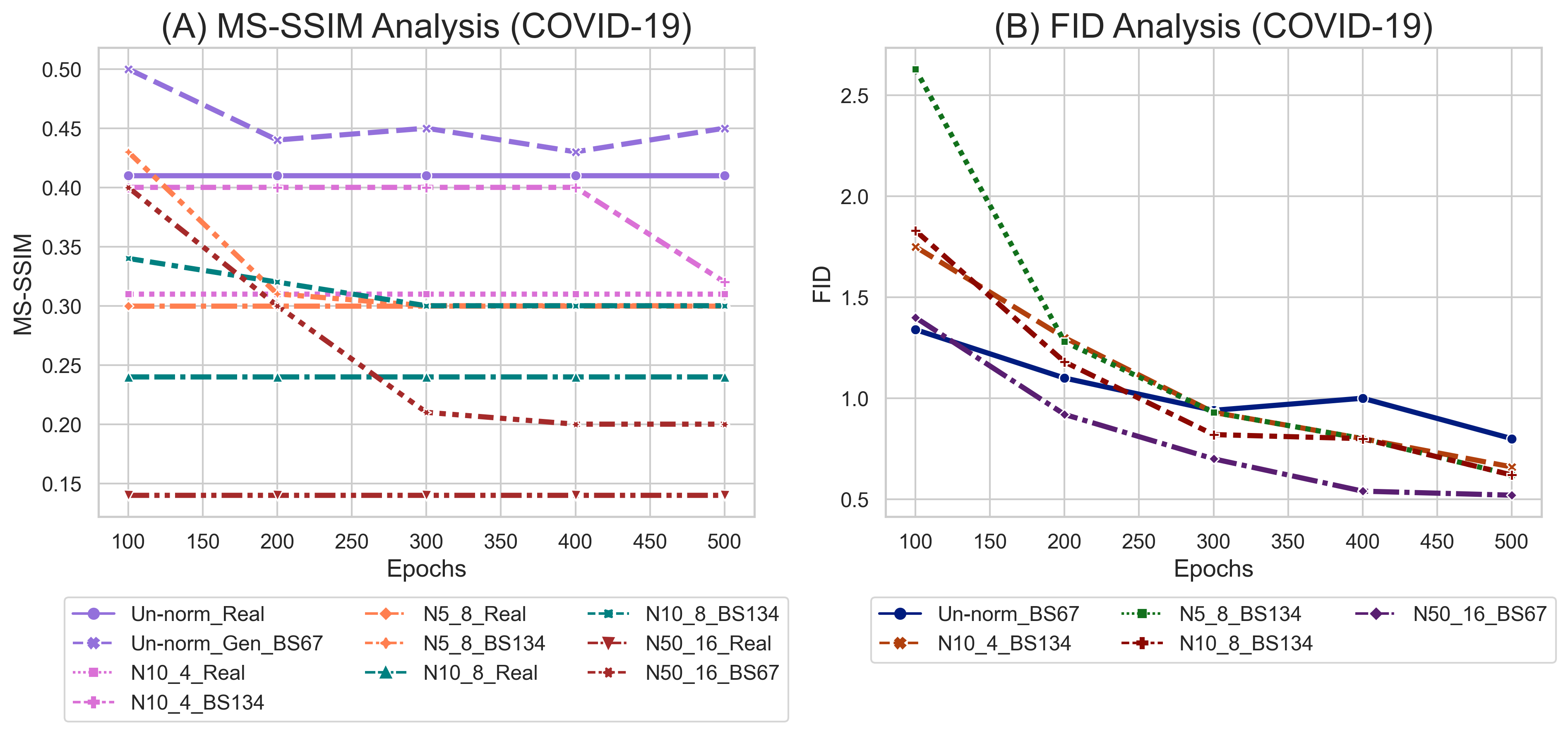}%
}

\subfloat[DCGAN Training behavior on PNEUMONIA X-ray images over epochs. The MS-SSIM scores of real and generated images are shown in the same colors.]{%
  \includegraphics[clip,width=1\textwidth]{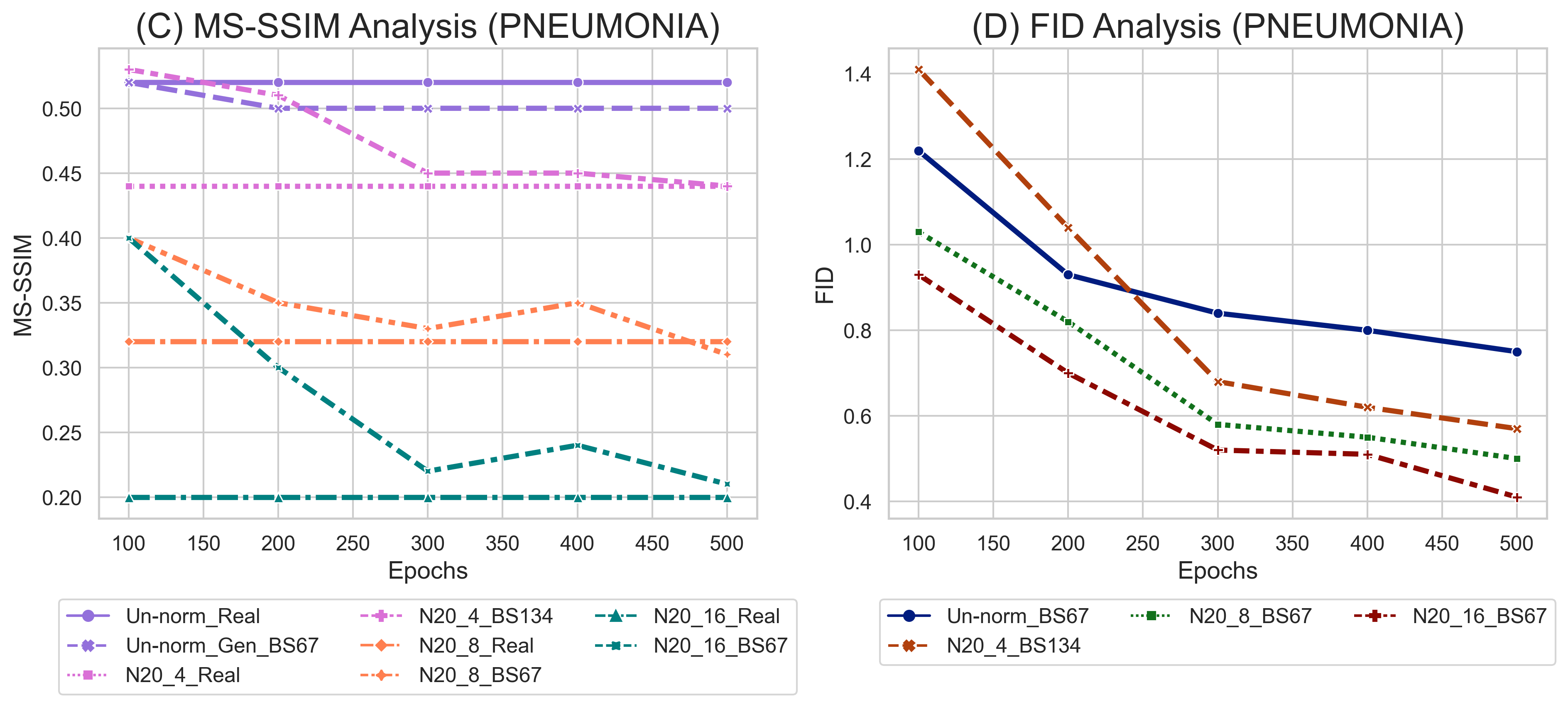}%
}
  \caption{DCGAN training behavior on Datasets A and B}
  \label{line_plots}
  \vspace{-1em}
\end{figure*}

\begin{figure*}[hbt!]
  \vspace{-1em}
  \centering
\subfloat[ACGAN Training behavior on COVID-19 vs Healthy X-ray images over epochs. The IS scores of real and generated images are shown in the same colors.]{%
  \includegraphics[clip,width=1\textwidth]{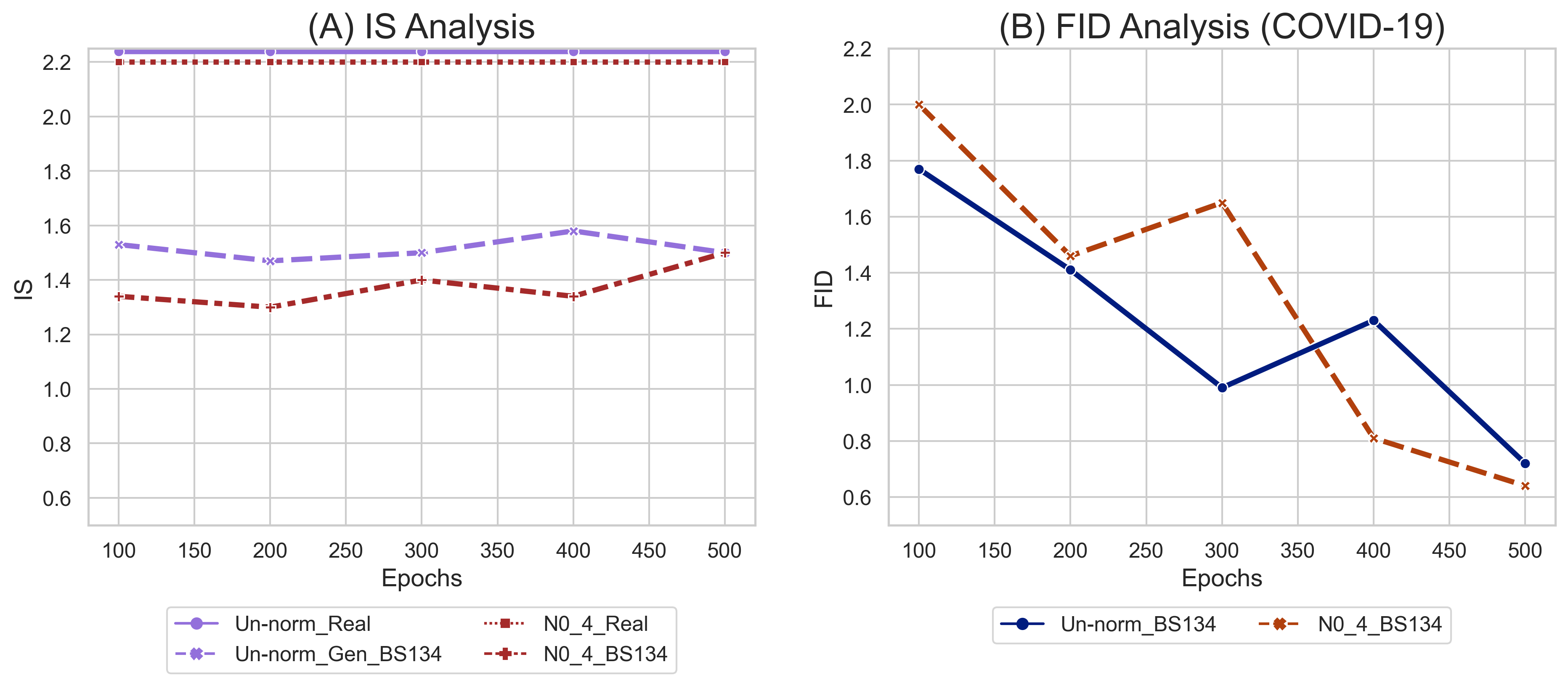}%
}

\subfloat[ACGAN Training behavior on PNEUMONIA vs Healthy X-ray images over epochs. The IS scores of real and generated images are shown in the same colors.]{%
  \includegraphics[clip,width=1\textwidth]{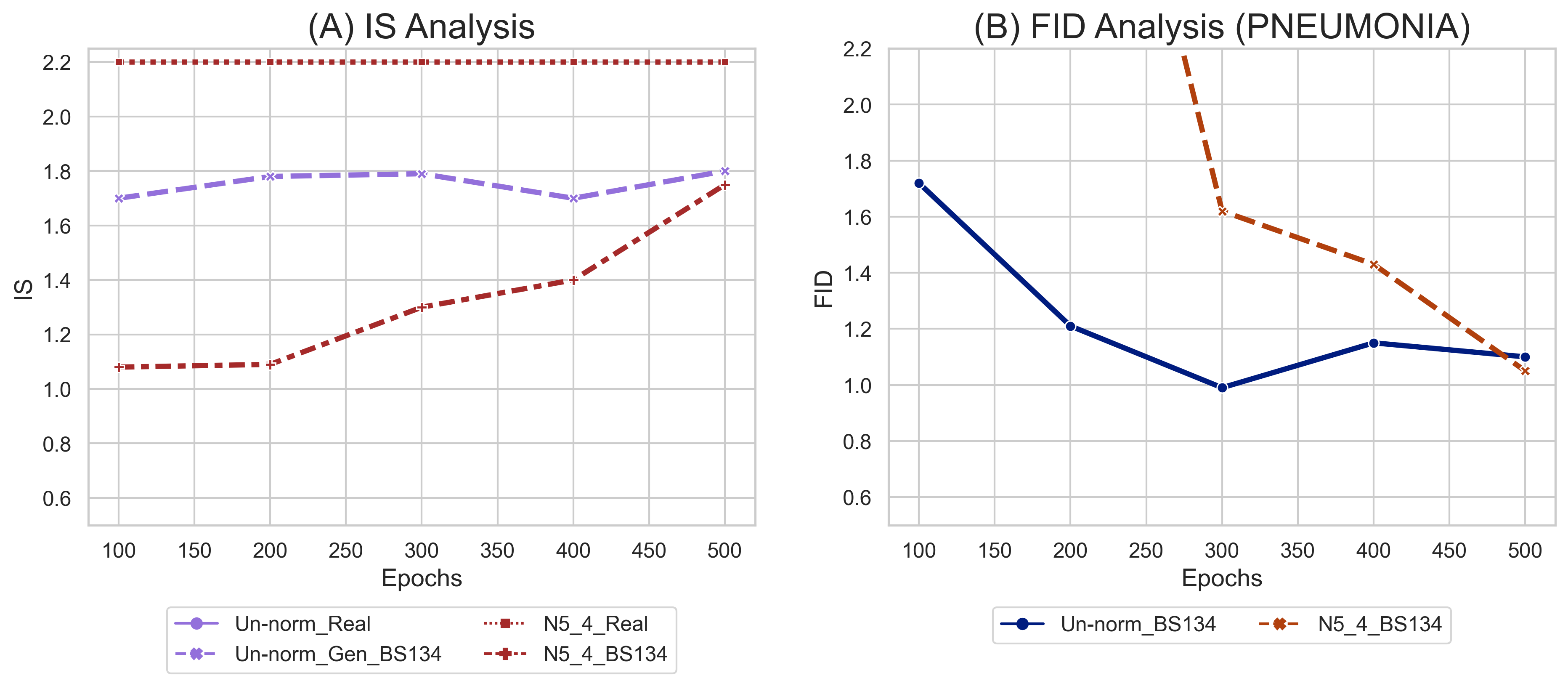}%
}
  \caption{ACGAN training behavior on Datasets A and B}
  \label{acganline_plots}
  \vspace{-1em}
\end{figure*}

In this paper, the intra-class mode collapse in the DCGAN is identified by the higher MS-SSIM score of un-normalized synthetic X-ray images than real images. The AIIN has alleviated the intra-class mode collapse by improving the capacity of DCGAN to generate improved diversified normalized synthetic images as evidenced by the improved MS-SSIM scores of generated X-ray images. The training behavior of DCGAN for the AIIN normalized over un-normalized images is also improved to generate diversified images as indicated in Fig. \ref{line_plots}. Figure \ref{line_plots} shows the training behavior of DCGAN for the best-diversified scores of MS-SSIM of un-normalized and AIIN-normalized X-ray images. The inter-class mode collapse in the ACGAN is identified by a lower IS score for synthetic images than for real images. The AIIN has alleviated the inter-class mode collapse by improving the ACGAN's capacity to generate diversified synthetic images, as evidenced by the improved IS scores of synthetic images. Figure \ref{acganline_plots} shows the training behavior of the ACGAN for the best-diversified IS scores of un-normalized and normalized X-ray images.

The intra-class diversity of synthetic images generated by the DCGAN and ACGAN is also improved using the AIIN normalized X-ray images as shown by the improved FID scores. FID analysis indicates the efficacy of using AIIN with GANs in improving the diversity of synthetic images. For the ACGAN, the IS and FID analyses show that the AIIN has a limited impact on alleviating the inter-class mode collapse using the AIIN for generating diversified synthetic X-ray images. The discriminator of the ACGAN finds it difficult to classify synthetic images from real images of binary classes. The AIIN has normalized the images of both classes where features of X-ray images of one class resemble the X-ray image features of the other class. So, the discriminator sends similar gradient feedback to the generator, which generates identical images repeatedly.

\begin{figure*}[hbt!]
  \vspace{-1em}
  \centering
\subfloat[IS scores enabling an assessment of the occurrence of inter-class mode collapse for Healthy and COVID-19 (Dataset-A) X-ray images.]{%
  \includegraphics[clip,width=0.95\textwidth,height=5.5cm]{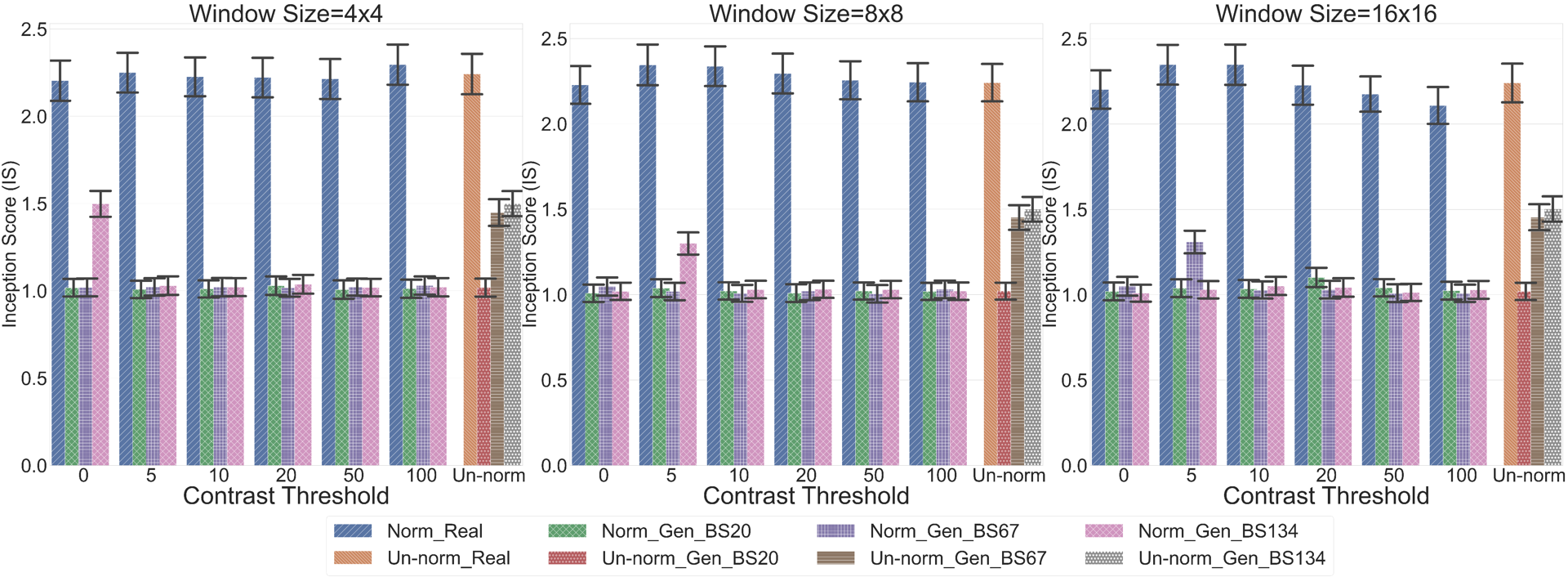}%
}

\subfloat[FID scores enabling an assessment of the intra-class diversity of COVID-19 X-ray images.]{%
  \includegraphics[clip,width=0.95\textwidth,height=5.5cm]{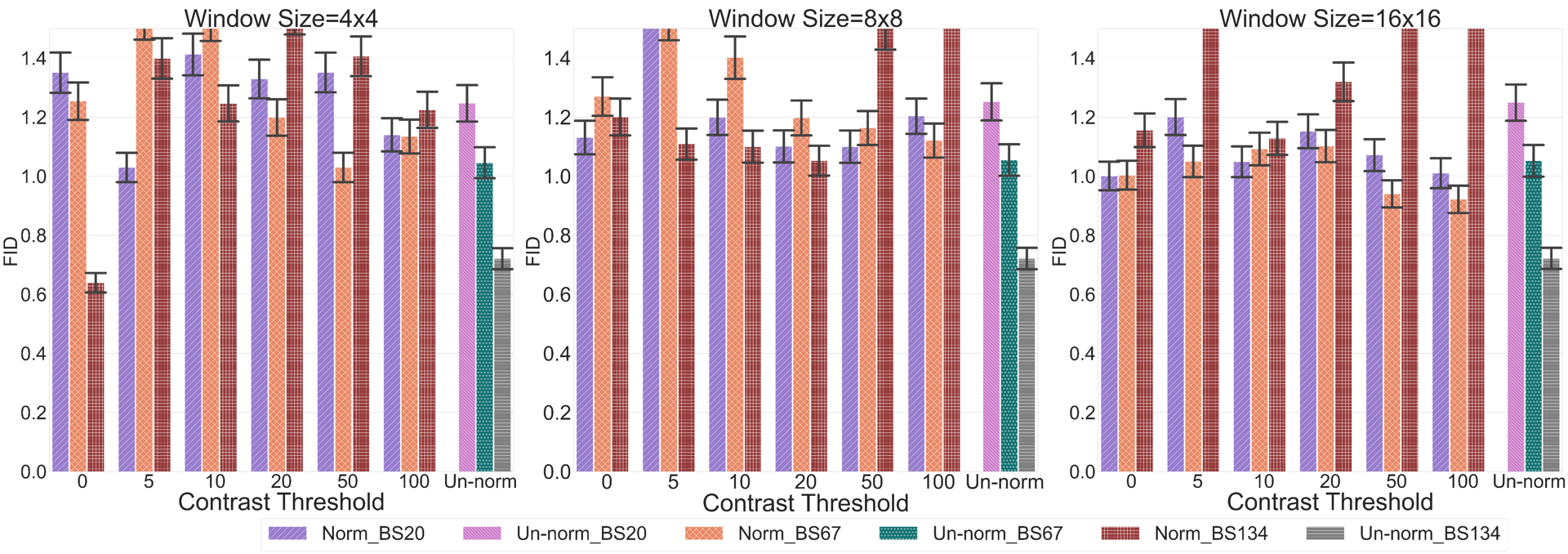}%
}
  \caption{The IS and FID scores of ACGAN for the un-normalized and AIIN normalized healthy and COVID-19 (Dataset-A) X-ray images.}
  \label{IS_FID_covid}
  \vspace{-1em}
\end{figure*}

\begin{figure*}[hbt!]
  \vspace{-1em}
  \centering
\subfloat[IS scores enabling an assessment of the occurrence of inter-class mode collapse for Healthy and Pneumonia (Dataset-B) X-ray images.]{%
  \includegraphics[clip,width=0.95\textwidth,height=5.5cm]{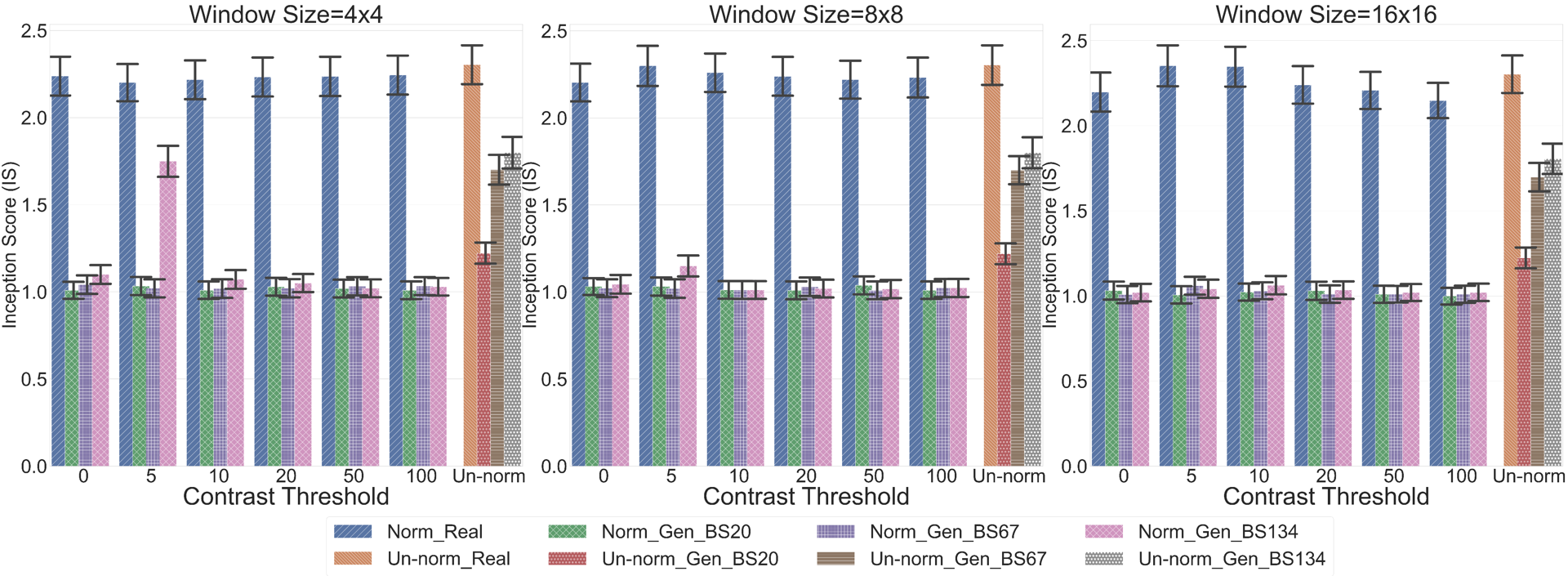}%
}

\subfloat[FID scores enabling an assessment of the intra-class diversity of Pneumonia X-ray images.]{%
  \includegraphics[clip,width=0.95\textwidth,height=5.5cm]{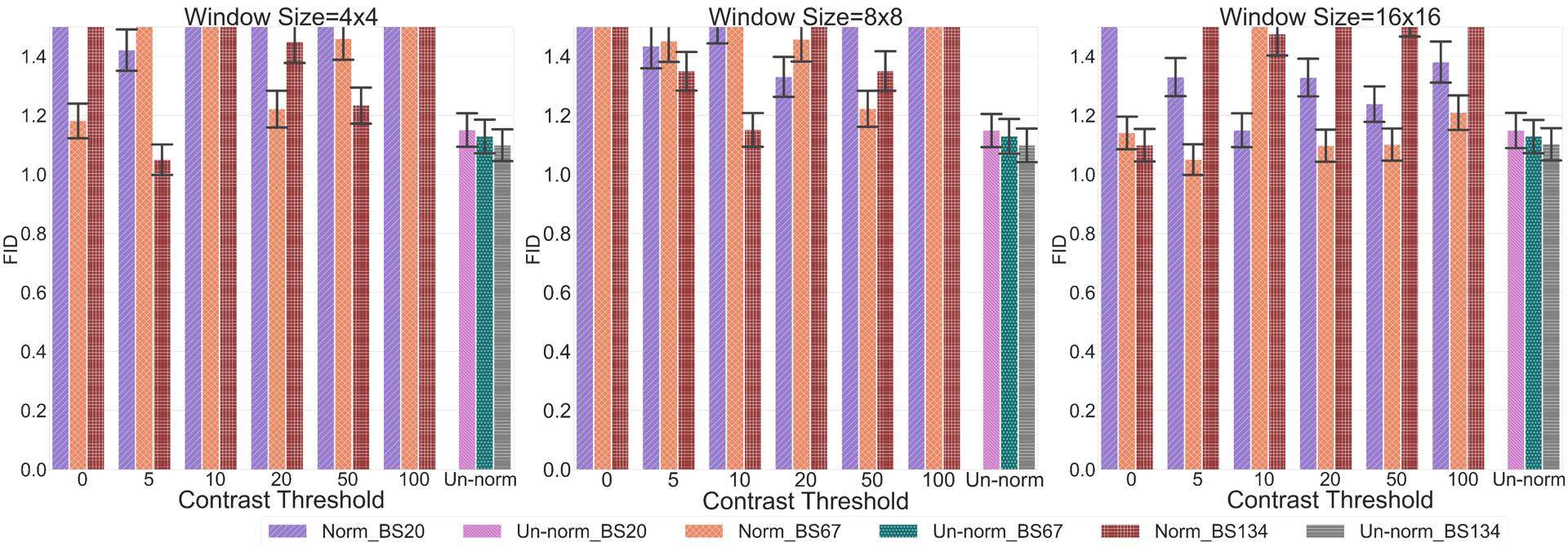}%
}
  \caption{The IS and FID scores of ACGAN for the un-normalized and AIIN normalized healthy and Pneumonia (Dataset-B) X-ray images.}
  \label{IS_FID_pneu}
  \vspace{-1em}
\end{figure*}

Parameters such as batch size, window size, and contrast threshold have substantially impacted the generation of diversified X-ray images as indicated by the varying MS-SSIM, FID, and IS scores. The batch size of 20 degrades the training performances of both DCGAN and ACGAN as compared to the larger batch sizes for un-normalized and normalized X-ray images for both datasets A and B, indicating the occurrence of mode collapse with lower diversity scores as shown in Fig. \ref{MS-SSIM_FID_covid}, \ref{MS-SSIM_FID_pneu}, \ref{IS_FID_covid}, and \ref{IS_FID_pneu}. In DCGAN and ACGAN, a batch size of 20 limits the ability of the generator to capture the full range of diverse features in X-ray images during each iteration of the training of GANs. It also limits the ability of the discriminator to provide less accurate estimations of the gradients that lead the generator towards mode collapse problems. However, AIIN has also reduced the effects of mode collapse in DCGAN and ACGAN with batch sizes of 20, but still, the synthetic images have degraded scores as compared to larger batch sizes.

\begin{table*}[hbt!]
\vspace{-1em}
\centering
\caption{ViT's classification performance of COVID-19 vs Healthy and Pneumonia vs Healthy chest X-ray images with different GAN-based scenarios. MS-SSIM evaluates the intra-class mode collapse in DCGAN while IS evaluates the inter-class mode collapse in ACGAN. Both scores are achieved by subtracting the scores of real datasets from the scores of generated datasets. Bold values show the best performances of the ViT classifier achieved with AIIN-DCGAN and AIIN-ACGAN augmented datasets.}
\begin{tabular}{p{0.5cm}p{1.8cm}p{1.5cm}p{1cm}p{0.5cm}p{1.8cm}p{0.7cm}p{1.2cm}p{1.3cm}p{0.8cm}p{1.4cm}}
       \toprule
       \textbf{Clf} & \textbf{Augmentation} & \textbf{Dataset} & \textbf{WS} & \textbf{CT} & \textbf{\textit{IS}/MS-SSIM} & \textbf{FID} & \textbf{Accuracy (\%)} & \textbf{Precision} & \textbf{Recall} & \textbf{F1-score} \\ \midrule
       ViT & x & COVID-19 & x & x & x & x & 86.0 & 0.85 & 0.72 & 0.78 \\
       ViT & DCG\_BS67 & COVID-19 & x & x & +0.04 & 0.78 & 86.0 & 0.84 & 0.70 & 0.76 \\
       ViT & DCG\_NBS134 & COVID-19 & 4x4 & 10 & +0.01 & 0.66 & 60.0 & 0.35 & 0.23 & 0.28 \\ 
       ViT & DCG\_NBS134 & COVID-19 & 8x8 & 5 & +0.02 & 0.62 & 55.0 & 0.27 & 0.20 & 0.23 \\
       ViT & DCG\_NBS134 & COVID-19 & 8x8 & 10 & +0.03 & 0.62 & 52.0 & 0.27 & 0.25 & 0.26 \\
       ViT & DCG\_NBS67 & COVID-19 & 16x16 & 50 & +0.04 & 0.52 & 38.0 & 0.02 & 0.02 & 0.02 \\
       ViT & ACG\_BS134 & COVID-19 & x & x & -0.74* & 0.72 & 85.0 & 0.83 & 0.69 & 0.76 \\
       ViT & ACG\_NBS134 & COVID-19 & 4x4 & 0 & -0.70* & 0.64 & 77.0 & 0.62 & 0.79 & 0.69 \\
       ViT & x & Pneumonia & x & x & x & x & 98.0 & 1.00 & 0.94 & 0.97 \\
       ViT & DCG\_BS67 & Pneumonia & x & x & -0.02 & 0.75 & 98.0 & 1.00 & 0.95 & 0.97 \\
       ViT & DCG\_NBS134 & Pneumonia & 4x4 & 20 & 0.00 & 0.57 & \textbf{98.0} & \textbf{0.96} & \textbf{0.97} & \textbf{0.96} \\
       ViT & DCG\_NBS67 & Pneumonia & 8x8 & 20 & -0.01 & 0.50 & 95.0 & 0.92 & 0.94 & 0.93 \\
       ViT & DCG\_NBS67 & Pneumonia & 16x16 & 20 & +0.01 & 0.41 & 94.0 & 0.88 & 0.93 & 0.91 \\
       ViT & ACG\_BS134 & Pneumonia & x & x & -0.5* & 1.10 & \textbf{98.0} & \textbf{0.97} & \textbf{0.97} & \textbf{0.97} \\
       ViT & ACG\_NBS134 & Pneumonia & 4x4 & 5 & -0.45* & 1.05 & 94.0 & 0.95 & 0.87 & 0.91 \\
       \bottomrule
       \multicolumn{11}{l}{Clf: Classifier; ViT: Vision Transformer; DCG: DCGAN; N: Normalized data; BS: Batch Size; CT: Contrast Threshold} \\
       \multicolumn{11}{l}{WS: Window Size; (*) symbol refers to the IS scores}
       \end{tabular}
       \label{tabclasscompare}
       \vspace{-1em}
\end{table*}

\begin{table*}[hbt!]
\vspace{-1em}
\centering
\caption{A comparison of the proposed methodology for the binary classification of COVID-19 vs Healthy and Pneumonia vs Healthy chest X-ray images using AIIN-DCGAN and AIIN-ACGAN-based augmentation to the prior state-of-the-art classification approaches. Datasets with augmented classes are highlighted in bold. The best classification score is also highlighted in bold values.}
\begin{footnotesize}
      \begin{tabular}{p{0.5cm}p{1cm}p{1.7cm}p{1.8cm}p{3cm}p{2.3cm}p{2cm}p{0.8cm}}
       \toprule
       \textbf{Ref} & \textbf{Dataset Source} & \textbf{Pre-Aug. dataset} & \textbf{Trad. Aug.} & \textbf{GANs Aug.} & \textbf{Post-Aug. dataset} & \textbf{Classifier} & \textbf{Acc. (\%)} \\ \midrule
       \cite{bodapati2022chxcapsnet} & \cite{kermany2018} & Healthy:1583, Pneumonia:4275 & Flip, Shear, Zoom & x & Missing & InceptionV3Caps & 94.84 \\
       \cite{krishnan2021visionT} & \cite{rahman2021exploring} & Healthy:9443, COVID:4831 & Br., Cont., Flip, Rot. & x & Healthy:9443, \textbf{COVID:9662} & ViT-B/32 & 97.6 \\
       \cite{manickam2021automated} & \cite{kermany2018} & Healthy:1346, Pneumonia:3883 & Gaussian\_Blur, Flip, Rot., Shift & x & \textbf{Healthy:4266}, Pneumonia:3883 & ResNet50 & 93.06 \\
       \cite{saad2022addressing} & \cite{kermany2018} & Healthy:1340, Pneumonia:3875 & Rot., Shear, Zoom & DCGAN(Healthy:1340) & \textbf{Healthy:2680}, Pneumonia:3875 & CNN & 91.50 \\
       This Work & \cite{rahman2021exploring} & Healthy:2680, Pneumonia:1340 & x & DCGAN(Pneumonia:1340) & Healthy:2680, \textbf{Pneumonia:2680} & ViT & \textbf{98.0} \\
       This Work & \cite{rahman2021exploring} & Healthy:2680, Pneumonia:1340 & x & ACGAN(Pneumonia:1340) & Healthy:2680, \textbf{Pneumonia:2680} & ViT & \textbf{98.0} \\
       \bottomrule
      \multicolumn{8}{l}{Ref: Reference; Trad: Traditional; Aug: Augmentation; ViT: Vision Transformer; Cont: Contrast; Br: Brightness; Rot: Rotation} \\
      \multicolumn{8}{l}{Tran: Translation; Acc: Accuracy}
       \end{tabular}
       \end{footnotesize}
       \label{tabcompare_pr-work}
       \vspace{-1em}
\end{table*}
In this paper, the results of ViT are improved for Pneumonia while degraded for COVID-19 X-ray images using AIIN-normalized augmented datasets. Results are evaluated using higher values of accuracy, precision, recall, and F1 scores for the targeted class being augmented. For COVID-19 images, the classification results of ViT are degraded because COVID-19 images contain suppressed organs such as bone and lung segments that are difficult to detect by the AIIN technique. As a result, noisy features are introduced as diverse features in these images during contrast-based normalization. For Pneumonia images, the precision, recall, and F1 scores for the contrast threshold of 20 with a window size 4x4  are improved because these images contain dense bones and lung features with low visibility in the images that can be detected by AIIN. The higher variation in the precision, recall, and F1 scores indicates that the classifier is not performing well on the training images thereby predicting the lower scores for test images. However, improving the classification metrics such as precision, recall, and F1 scores collectively indicates that the classifier captures true positives and false positives of AIIN normalized image samples accurately. The classification scores are also compared with the prior state-of-the-art classification approaches of COVID-19 and Pneumonia X-ray images as shown in Table \ref{tabcompare_pr-work}. Table \ref{tabcompare_pr-work} indicates that the efficacy of the AIIN with the DCGAN and the ACGAN improved the classification scores that outperformed the prior work. Furthermore, the comparative analysis in Table \ref{tabcompare_pr-work} also indicates that this work improves classification scores with GAN-base synthetic images without using traditional augmentation approaches demonstrating the efficacy of synthetic images.

\subsection{Limitations of this work}
In this work, AIIN is implemented with the GAN architectures such as DCGAN for addressing intra-class mode collapse and ACGAN for addressing inter-class mode collapse problems. AIIN has shown a capacity to improve the intra-class and inter-class diversities of synthetic images generated by DCGAN and ACGAN. However, AIIN can be utilized with advanced GAN architectures such as StyleGAN2 \cite{karras2019style}, StyleGAN3 \cite{karras2021alias}, and StyleSwin \cite{zhang2022styleswin} to improve the generation of diversified and high-quality synthetic images in the biomedical imaging domain.

This work is limited to X-ray imaging modalities. AIIN technique can improve the generation of alternative biomedical imaging modalities as well such as dermoscopic images, retinal fundus images, histopathology images, etc.

FID and IS metrics have been widely utilized to assess the diversity and quality of synthetic images in the domain of biomedical imaging \cite{abdelhalim2021data} \cite{qin2020gan} \cite{huang2020enhanced} \cite{han2020breaking}. These metrics use the Inception-Net pre-trained on the ImageNet \cite{deng2009imagenet} dataset. The ImageNet \cite{deng2009imagenet} dataset has no class containing biomedical images. Therefore, FID and IS can be biased in evaluating the diversity of synthetic images. It is an open research problem to investigate a suitable metric for domain-specific imagery in a biomedical setting.

\section{Conclusion}
\label{sec:conclusion}
In this work, a preprocessing technique AIIN is proposed to alleviate the intra-class mode collapse in DCGAN and inter-class mode collapse in ACGAN architectures for X-ray image synthesis. Results show that the AIIN has alleviated the mode collapse problems in both types of GANs and helps to generate improved diversified synthetic X-ray images as evidenced by the improved MS-SSIM and FID scores.

Furthermore, synthetic images are utilized to augment the imbalanced X-ray image datasets. This work concludes that the AIIN in GANs significantly relies on the type of modality that requires careful attention in selecting suitable window size and contrast threshold values. The synthetic images generated by the AIIN-based GANs can better augment the imbalanced datasets and help to achieve higher classification scores.

\printcredits

\section*{Declaration of interest statement}
None declared.

\section*{Statements of ethical approval}
No ethical approval is required.

\section*{Acknowledgements}
This research work is supported by the Risam scholarship award offered by Munster Technological University Cork, Ireland.

%


\end{document}